\documentstyle[preprint,aps,eqsecnum]{revtex}
\tighten
\begin{document}
\draft
\hyphenation{Rijken}
\hyphenation{Nijmegen}

\title{Soft two-meson-exchange nucleon-nucleon potentials.
       \\ II. One-pair and two-pair diagrams}

\author{Th.A.\ Rijken}
\address{Institute for Theoretical Physics, University of Nijmegen,
         Nijmegen, The Netherlands}

\author{V.G.J.\ Stoks\footnote{Present address:
        TRIUMF, 4004 Wesbrook Mall, Vancouver, British Columbia,
        Canada V6T 2A3}}
\address{Department of Physics,
         The Flinders University of South Australia,
         Bedford Park, South Australia 5042, Australia}

\date{version of: \today}
\maketitle

\begin{abstract}
Two-meson-exchange nucleon-nucleon potentials are derived where
either one or both nucleons contains a pair vertex.
Physically, the meson-pair vertices are meant to describe in an
effective way (part of) the effects of heavy-meson exchange and
meson-nucleon resonances. {}From the point of view of ``duality,''
these two kinds of contribution are roughly equivalent.
The various possibilities for meson pairs coupling to the nucleon
are inspired by the chiral-invariant phenomenological Lagrangians
that have appeared in the literature. The coupling constants are
fixed using the linear $\sigma$ model. We show that the inclusion
of these two-meson exchanges gives a significant improvement over
a potential model including only the standard one-boson exchanges.
\end{abstract}
\pacs{13.75.Cs, 12.39.Pn, 21.30.+y}

\narrowtext

\section{INTRODUCTION}
\label{sec:chap1}
In this second paper on two-meson-exchange potentials, we derive the
contributions to the nucleon-nucleon potentials when either one or
both nucleons contains a pair vertex. The corresponding ``seagull''
diagrams are labeled as one-pair and two-pair diagrams, and are
depicted in Fig.~\ref{pap2fig1}. They are in a different class than
the planar and crossed-box diagrams. The latter can be understood as
the second-order contributions in a series expansion of multi-meson
exchanges and were calculated in the previous paper~\cite{Rij95a}.
The two types of two-meson-exchange potentials presented here and in
the previous paper~\cite{Rij95a} are part of our program to study the
effect of two-meson exchanges in potential models, and to extend the
Nijmegen soft-core one-boson-exchange potential~\cite{Nag78} to arrive
at a new extended soft-core nucleon-nucleon model, hereafter referred
to as the ESC potential.

Our motivation for deriving the pair-meson potentials comes from
``duality''~\cite{Dol68,Swa89}. Since we do not explicitly include any
contributions to the potentials coming from intermediate states with
nucleon resonances, we have indicated in Fig.~\ref{pap2fig2} how these
resonances can be included via meson-pair contributions.
According to ``duality,'' the resonance contributions to the various
meson-nucleon amplitudes can be described approximately by heavy-meson
exchanges. Treating the heavy-meson propagators as constants, which
should be adequate at low energies, leads directly to pair-meson
exchanges.
Hence, the pair-meson potentials can be viewed as the result of
integrating out the heavy-meson and resonance degrees of freedom. We
are less radical than Weinberg (see, e.g., Ref.~\cite{Wei90}), in that
we do not integrate out the degrees of freedom of the mesons with
masses below 1 GeV.

The pairs we consider are $\pi\pi$, $\pi\rho$, $\pi\varepsilon$, and
$\varepsilon\varepsilon$. They are inspired by the chiral-invariant
phenomenological Lagrangians that have appeared in the literature,
see for example \cite{Sch68,Wei68}.
Here the $\varepsilon(760)$ scalar meson is treated as a very broad
meson with a width of $\Gamma_{\varepsilon}\sim640$ MeV~\cite{remark}.
The potential due to the exchange of a broad meson can be approximated
by the sum of two potentials where each potential is due to the exchange
of a stable meson~\cite{Sch71,Bin71}. Because of the large width of the
$\varepsilon$(760), the low-mass pole in the two-pole approximation is
rather small ($\sim500$ MeV). Hence, it is expected to contribute
significantly to the two-meson potentials, which is the reason why we
explicitly included the $\varepsilon\varepsilon$ pair potential.
To emphasize the importance of the low-mass contribution, we will
henceforth denote these scalar contributions to the pair potentials by
$\pi\sigma$ and $\sigma\sigma$, where $\sigma$ stands for the low-mass
pole.

We neglect the contributions from the negative-energy states of the
nucleons. We assume that at low energies a strong ``nucleon-pair
suppression'' mechanism (i.e., suppression of the nucleon-antinucleon
$N\!\overline{N}$ pairs) is operative. This is due to the compositeness
and the large mass of the nucleons. Obviously, this is only valid in
the low-energy regime. {}From relativity we know that nucleon-pair
suppression cannot be absolute. However, we mention that it is very
plausible from the point of view~\cite{Swa78} of the nonrelativistic
quark model (NRQM).
Also, in the chiral quark-model picture~\cite{Man84} it is rather
unlikely that $N\!\overline{N}$ pairs play a role in the low-energy
region. Moreover, nucleon-pair suppression is substantiated by studies
of the large $N_{c}$ limit in QCD~\cite{Wit79}.
In this case then, the Thomson limit is provided by the negative-energy
contributions of the constituents, in particular the current quarks.
A covariant description of this nucleon-pair suppression could very
well be done by introducing off-mass-shell factors to the
meson-nucleon-nucleon vertices. Instead of attempting such a covariant
description in this paper, we simply neglect the transitions to the
negative-energy states.

The paper is organized as follows.
In Sec.~\ref{sec:chap2} the meson-pair-exchange kernels are derived.
We give the interaction Hamiltonians and the vertices in Pauli-spinor
space, while the implementation of the Gaussian form factors is briefly
indicated. In Sec.~\ref{sec:chap3} we derive the one-pair and two-pair
potentials. We then extend the calculations by including the $1/M$
corrections from the pseudovector vertex and from the nonadiabatic
expansion of the energy denominators.
The strength of the pair interactions can be estimated using the
saturation with one-boson exchanges, but this involves some ambiguity
in how to approximate the intermediate-boson propagator.
In Sec.~\ref{sec:chap5} we therefore choose to fix the pair coupling
constants at their values as obtained in the linear $\sigma$
model~\cite{Gel60}. These values do not have to be taken exactly,
of course, but they clearly suffice to demonstrate the improvements
that can be obtained when we include these pair interactions in a
nucleon-nucleon potential model.
Plots of the various pair potentials are then shown and discussed in
Sec.~\ref{sec:chap6}. In this section we also compare this new ESC
potential with the one-boson-exchange Nijm93 potential~\cite{Sto94},
demonstrating the significant improvement in the description of the
scattering data that has been obtained by including the two-meson
exchanges.

Finally, some details on parts of the calculation are collected in
two appendices, while the coordinate-space pair potentials are given
explicitly in Appendix~\ref{app:appC}.

\section{THE PAIR-MESON-EXCHANGE KERNEL}
\label{sec:chap2}
The fourth-order two-meson-exchange kernel is derived following the
procedure as discussed in the previous paper~\cite{Rij95a}, which
closely follows an earlier publication~\cite{Rij91}, where we derived
the soft two-pion-exchange potential. For details and definitions we
refer to these two references.

Rather than repeating the derivation again, we here immediately proceed
to give the one-pair and two-pair kernels.
Introducing subscripts $1P$ and $2P$, the one-pair kernel is given by
\widetext
\begin{eqnarray}
  K_{1P}({\bf p}',{\bf p}|W)_{a'b';ab}&=&-(2\pi)^{-2}
         [W-{\cal W}({\bf p}')]\;[W-{\cal W}({\bf p})]
\nonumber\\ &\times&
  \sum_{a'',b''}\int\!dp_{0}'\int\!dp_{0}\int\!dk_{10}\int\!dk_{20}
                \int\!d{\bf k}_{1}\int\!d{\bf k}_{2}      %\nonumber\\
          \ i(2\pi)^{-4}\delta^{4}(p'-p-k_{1}-k_{2})
\nonumber\\ &\times&
  [k^{2}_{2}-m^{2}_{2}+i\delta]^{-1}\left[F^{(a')}_{W}({\bf p}',p_{0}')
     F^{(b')}_{W}(-{\bf p}',-p_{0}') \right]^{-1}          \nonumber\\
  &\times& \left\{ [\Gamma_{j,i}({\bf p}',p'_{0};{\bf p},p_{0})]
        [\Gamma_{j} F_{W}^{-1}(-{\bf p}-{\bf k}_{1},-p_{0}-k_{10})
         \Gamma_{i}]^{(b'')}    \right.                    \nonumber\\
  && + \left. [\Gamma_{j} F_{W}^{-1}
        ({\bf p}+{\bf k}_{1}, p_{0}+k_{10}) \Gamma_{i}]^{(a'')}
        [\Gamma_{j,i}(-{\bf p}',-p'_{0};-{\bf p},-p_{0})]\right\}
                                                          \nonumber\\
  &\times& \left[F^{(a)}_{W}({\bf p},p_{0})F^{(b)}_{W}
      (-{\bf p},-p_{0})\right]^{-1}[k^{2}_{1}-m^{2}_{1}+i\delta]^{-1},
                                        \label{ker1P}
\end{eqnarray}
whereas the two-pair kernel is given by
\begin{eqnarray}
  K_{2P}({\bf p}',{\bf p}|W)_{a'b';ab}&=&-(2\pi)^{-2}
         [W-{\cal W}({\bf p}')]\;[W-{\cal W}({\bf p})]
\nonumber\\ &\times&
                \int\!dp_{0}'\int\!dp_{0}\int\!dk_{10}\int\!dk_{20}
                \int\!d{\bf k}_{1}\int\!d{\bf k}_{2}      %\nonumber\\
         \ i(2\pi)^{-4}\delta^{4}(p'-p-k_{1}-k_{2})
\nonumber\\ &\times&
  [k^{2}_{2}-m^{2}_{2}+i\delta]^{-1}\left[F^{(a')}_{W}({\bf p}',p_{0}')
     F^{(b')}_{W}(-{\bf p}',-p_{0}') \right]^{-1}          \nonumber\\
  &\times& \left\{ [\Gamma_{j,i}({\bf p}',p'_{0};{\bf p},p_{0})]
               [\Gamma_{j,i}(-{\bf p}',-p'_{0};-{\bf p},-p_{0})]
           \right\}                                        \nonumber\\
  &\times& \left[F^{(a)}_{W}({\bf p},p_{0})F^{(b)}_{W}
     (-{\bf p},-p_{0})\right]^{-1}[k^{2}_{1}-m^{2}_{1}+i\delta]^{-1}.
                                        \label{ker2P}
\end{eqnarray}
\narrowtext
Here $m_{1}$ and $m_{2}$ denote the two meson masses, $\Gamma_{i}$ and
$\Gamma_{j}$ denote the nucleon-nucleon-meson vertices, and
$\Gamma_{j,i}$ denotes the nucleon-nucleon-meson-meson vertex;
they follow from the interaction Hamiltonians (see below).
Because we only consider nucleons in the intermediate state, we have
$a=a'=a''=N$ and $b=b'=b''=N$.
Note that the first term between the curly brackets in the one-pair
kernel corresponds to having the pair vertex on the first nucleon
(hence, no label $a''$), and the second term to having the pair vertex
on the other nucleon (no label $b''$).

{}From the explicit expressions (\ref{ker1P}) and (\ref{ker2P}), it
is clear that one can perform the integration over the energy
variables $p'_{0}$, $p_{0}$, $k_{10}$, and $k_{20}$. The execution
of these integrals is quite similar to those worked-out explicitly
in Ref.~\cite{Rij91}, and so are the results. Details are given
in Appendix~\ref{app:appA}. In the case of the one-pair diagrams,
we have taken the diagram where the meson-pair vertex is on line $a$.

In the adiabatic approximation, i.e., $E({\bf p})\approx M$, the
energy denominators of the various time-ordered diagrams are
\begin{eqnarray}
  D^{(1)}_{a}(\omega_{1},\omega_{2}) &=& \frac{1}{2\omega_{1}\omega_{2}}
        \,\frac{1}{\omega_{2}(\omega_{1}+\omega_{2})},   \nonumber\\
  D^{(1)}_{b}(\omega_{1},\omega_{2}) &=&
          \frac{1}{2\omega_{1}^{2}\omega_{2}^{2}},       \nonumber\\
  D^{(1)}_{c}(\omega_{1},\omega_{2}) &=& \frac{1}{2\omega_{1}\omega_{2}}
        \,\frac{1}{\omega_{1}(\omega_{1}+\omega_{2})},   \nonumber\\
  D^{(2)}(\omega_{1},\omega_{2}) &=& -\frac{1}{2\omega_{1}\omega_{2}}
        \,\frac{1}{\omega_{1}+\omega_{2}}.             \label{Dpairs}
\end{eqnarray}
Here we have labeled the three time-ordered one-pair graphs by
$a,b,c$, which correspond, respectively, to the diagrams depicted in
Fig.~\ref{pap2fig3}. However, since we always have to add the
contribution from the ``mirror'' graphs anyway (i.e., with the pair
vertex at line $b$), we have already included the resulting factor
of 2 in the $D^{(1)}$ energy denominators of Eq.~(\ref{Dpairs}).
The two time-ordered graphs for the two-pair diagrams are each
other's ``mirror'' graphs. In most cases the vertices do not explicitly
depend on the time ordering and we can add the three time-ordered
one-pair energy denominators to give
\begin{equation}
   D^{(1)}(\omega_{1},\omega_{2}) =
       D^{(1)}_{a} + D^{(1)}_{b} + D^{(1)}_{c} =
     \frac{1}{\omega_{1}^{2}\omega_{2}^{2}}.       \label{D1pair}
\end{equation}

Before we can proceed and calculate the pair-meson potentials, we
have to define the nucleon-nucleon-meson ($N\!Nm$) and
nucleon-nucleon-meson-meson ($N\!Nm_{1}m_{2}$) Hamiltonians. For
point couplings the nucleon-nucleon-meson Hamiltonians are
\begin{mathletters}
\begin{eqnarray}
  {\cal H}_{P} &=& \frac{f_{P}}{m_{\pi}}
     \bar{\psi} \gamma_{5}\gamma_{\mu}\bbox{\tau}\psi
     \!\cdot\!\partial^{\mu}\bbox{\phi}_{P},             \label{Lagpv}\\
  {\cal H}_{V} &=& g_{V}\bar{\psi}\gamma_{\mu}\bbox{\tau}
     \psi\!\cdot\!\bbox{\phi}^{\mu}_{V} - \frac{f_{V}}{2M}
     \bar{\psi}\sigma_{\mu\nu}\bbox{\tau}\psi\!\cdot\!\partial^{\nu}
     \bbox{\phi}^{\mu}_{V} \ ,                            \label{Lagv}\\
  {\cal H}_{S} &=& g_{S}\bar{\psi}\bbox{\tau}\psi
     \!\cdot\!\bbox{\phi}_{S} \ ,                         \label{Lags}
\end{eqnarray}
\end{mathletters}
where $\bbox{\phi}$ denotes the pseudovector-, vector-, and
scalar-meson field, respectively. For the isospin $I=0$ mesons, the
isospin Pauli matrices, $\bbox{\tau}$, are absent.

For the phenomenological meson-pair interactions, the Hamiltonians are
\begin{mathletters}
\begin{eqnarray}
  {\cal H}_{S} &=& \bar{\psi}\psi
       \left[g_{(\pi\pi)_{0}}\bbox{\pi}\!\cdot\!\bbox{\pi} +
           g_{(\sigma\sigma)}\sigma^{2}\right] / m_{\pi},  \label{LPs}\\
  {\cal H}_{V} &=& g_{(\pi\pi)_{1}}\bar{\psi}\gamma_{\mu}\bbox{\tau}\psi
       \cdot(\bbox{\pi}\!\times\!\partial^{\mu}\bbox{\pi})
       / m^{2}_{\pi}                                         \nonumber\\
  & &  -\frac{f_{(\pi\pi)_{1}}}{2M}\bar{\psi}\sigma_{\mu\nu}
       \bbox{\tau}\psi\partial^{\nu}\cdot(\bbox{\pi}\!\times\!
       \partial^{\mu}\bbox{\pi}) / m^{2}_{\pi},            \label{LPv}\\
  {\cal H}_{A} &=& g_{(\pi\rho)_{1}}
       \bar{\psi}\gamma_{5}\gamma_{\mu}\bbox{\tau}\psi\cdot
       (\bbox{\pi}\times\bbox{\rho}^{\mu}) / m_{\pi},      \label{LPa}\\
  {\cal H}_{P} &=& g_{(\pi\sigma)} \bar{\psi}\gamma_{5}\gamma_{\mu}
       \bbox{\tau}\psi \cdot (\bbox{\pi}\partial^{\mu}\sigma
       -\sigma\partial^{\mu}\bbox{\pi}) / m^{2}_{\pi}      \label{LPp}
\end{eqnarray}
\end{mathletters}

The transition from Dirac spinors to Pauli spinors is reviewed in
Appendix C of \cite{Rij91}. Following this reference and keeping only
terms up to order $1/M$, we find that the vertex operators in
Pauli-spinor space for the $N\!Nm$ vertices are given by
\widetext
\begin{mathletters}
\begin{eqnarray}
   \bar{u}({\bf p}')\Gamma^{(1)}_{P}u({\bf p}) &=&
     -i\frac{f_{P}}{m_{\pi}}\!\left[ \bbox{\sigma}_{1}\!\cdot\!{\bf k}
      \pm\frac{\omega}{2M}\bbox{\sigma}_{1}\!\cdot\!
               ({\bf p}'+{\bf p}) \right],            \label{Gam1p}\\
   \bar{u}({\bf p}')\Gamma^{(1)}_{V}u({\bf p}) &=&
     g_{V}\left[ \phi^{0}_{V} - \frac{1}{2M}\biggl\{({\bf p}'+{\bf p})
        +i(1+\kappa_{V})\bbox{\sigma}_{1}\!\times\!{\bf k}
          \biggr\}\!\cdot\!\bbox{\phi}_{V}\right],    \label{Gam1v}\\
   \bar{u}({\bf p}')\Gamma^{(1)}_{S}u({\bf p}) &=& g_{S}, \label{Gam1s}
\end{eqnarray}
\end{mathletters}
where we defined ${\bf k}={\bf p}'-{\bf p}$ and
$\kappa_{V}=f_{V}/g_{V}$.
In the pseudovector vertex, the upper (lower) sign stands for creation
(absorption) of the pion at the vertex.
Similarly, the $N\!Nm_{1}m_{2}$ vertices result in
\begin{mathletters}
\begin{eqnarray}
   \bar{u}({\bf p}')\Gamma^{(2)}_{S}u({\bf p}) &=&
       g_{(\pi\pi)_{0}}/m_{\pi} \ \ \ {\rm and}\ \ \
       g_{(\sigma\sigma)}/m_{\pi},                   \label{Gam2s}\\
   \bar{u}({\bf p}')\Gamma^{(2)}_{V}u({\bf p}) &=&
       ig_{(\pi\pi)_{1}}\left[(\pm\omega_{1}\mp\omega_{2})
        +\frac{1}{M}\biggl\{{\bf q}\!\cdot\!({\bf k}_{1}-{\bf k}_{2})
        -(1+\kappa_{1})\bbox{\sigma}_{1}\!\cdot\!
         ({\bf k}_{1}\times{\bf k}_{2})\biggr\}\right] / m^{2}_{\pi},
\nonumber\\&
                                                    \label{Gam2v}\\
   \bar{u}({\bf p}')\Gamma^{(2)}_{A}u({\bf p}) &=&
       g_{(\pi\rho)_{1}}\left[\bbox{\sigma}_{1}\!\cdot\!\bbox{\rho}
       -\frac{1}{M}\bbox{\sigma}_{1}\!\cdot\!{\bf q}\rho^{0}\right]
        / m_{\pi},                                  \label{Gam2a}\\
   \bar{u}({\bf p}')\Gamma^{(2)}_{P}u({\bf p}) &=&
       ig_{(\pi\sigma)}\left[\bbox{\sigma}_{1}\!\cdot\!
           ({\bf k}_{1}-{\bf k}_{2}) - \frac{1}{M}
         \bbox{\sigma}_{1}\!\cdot\!{\bf q}(\pm\omega_{1}\mp\omega_{2})
       \right] / m^{2}_{\pi},                       \label{Gam2p}
\end{eqnarray}
\end{mathletters}
\narrowtext\noindent
where ${\bf q}={\textstyle\frac{1}{2}}({\bf p}'+{\bf p})$ and
$\kappa_{1}=(f/g)_{(\pi\pi)_{1}}$.
Again, the upper (lower) sign in front of $\omega_{1}$ and $\omega_{2}$
refers to creation (absorption) of the meson at the vertex.
For both the $N\!Nm$ and $N\!Nm_{1}m_{2}$ vertices, the expressions
for $\bar{u}(-{\bf p}')\Gamma u(-{\bf p})$ are trivially obtained by
substituting $({\bf p}',{\bf p},{\bf k}_{i},\omega_{i},\bbox{\sigma}_{1})
\rightarrow(-{\bf p}',-{\bf p},-{\bf k}_{i},\omega_{i},\bbox{\sigma}_{2})$.

The generalization of the interaction kernels to the case with a
Gaussian (or any other) form factor has been treated and explained
in \cite{Rij91}. We make the substitution
\begin{equation}
  [k^{2}-m^{2}+i\delta]^{-1} \longrightarrow
     \int_{0}^{\infty}\!d\mu^{2} \frac{\rho(\mu^{2})}
             {k^{2}-\mu^{2}+i\delta},
\end{equation}
for each meson-exchange line in the Feynman diagrams. Here,
$\rho(\mu^{2})$ is the spectral function, representing the form
factors involved in meson exchange. At low and medium energy, we
have to a very good approximation $t=k^{2}\approx-{\bf k}^{2}<0$,
and so for space-like momentum transfers we can use Gaussian form
factors $F({\bf k}^{2})=\exp(-{\bf k}^{2}/\Lambda^{2})$, where
$\Lambda$ denotes the cutoff mass. The Gaussian form factor is
introduced by the substitution
\begin{equation}
   \int_{0}^{\infty}\!d\mu^{2} \frac{\rho(\mu^{2})}
            {{\bf k}^{2}+\mu^{2}} \longrightarrow
   \frac{F({\bf k}^{2})}{{\bf k}^{2}+m^{2}}.
\end{equation}
The $N\!Nm$ and $N\!Nm_{1}m_{2}$ vertices have different
form factors. We will use
\begin{eqnarray}
  &&F_{N\!Nm_{1}m_{2}}({\bf k}_{1},{\bf k}_{2}) =
         \exp(-{\bf k}_{1}^{2}/2\Lambda_{1}^{2})\,
         \exp(-{\bf k}_{2}^{2}/2\Lambda_{2}^{2}),   \nonumber\\
  &&F_{N\!Nm}({\bf k}^{2}) = \exp(-{\bf k}^{2}/2\Lambda_{m}^{2}),
                                                  \label{formf}
\end{eqnarray}
where $\Lambda_{1}$ and $\Lambda_{2}$ are the form factor masses for
mesons $m_{1}$ and $m_{2}$, respectively. A motivation for this
prescription could be that in ``duality'' the structure of the
$N\!Nm_{1}m_{2}$ vertex is either saturated by heavy mesons or
meson-nucleon resonances. In this last case, assuming that the
meson-nucleon resonance transitions all have roughly the same
(inelastic) form factor, the form (\ref{formf}) is a natural one.

\section{MESON-PAIR POTENTIALS}
\label{sec:chap3}
In this paper we restrict ourselves mainly to the pion-meson pair
potentials. The only exception is the $\sigma\sigma$-pair potential,
but the derivation of the potential for any other combination of two
mesons will be straight forward. Here $\sigma$ stands for the low-mass
pole in the two-pole approximation~\cite{Sch71,Bin71} of the broad
$\varepsilon$(760) scalar meson.
(At this point it might be worthwhile to mention that in the previous
paper~\cite{Rij95a} we refrained from evaluating the $\sigma\sigma$
planar and crossed-box contributions. The reason for this is that the
leading-order contribution of a potential due to the exchange of two
isoscalar-scalar mesons is identically zero.)
In the description of the pion-meson pair potentials we always assign
the index 1 to the pion, and the index 2 to the other meson, which
can be a pion as well.

Armed with the Pauli-spinor vertices given in the previous section,
it is now straight forward to derive the one-pair and two-pair
potentials. They can be succinctly written as
\begin{eqnarray}
  V^{(n)}_{\rm pair}(\alpha\beta) &=&
     C^{(n)}(\alpha\beta) g^{(n)}(\alpha\beta)
     \int\!\!\int\!\frac{d^{3}k_{1}d^{3}k_{2}}{(2\pi)^{6}}   \nonumber\\
  && \times e^{i({\bf k}_{1}+{\bf k}_{2})\cdot{\bf r}}
       F_{\alpha}({\bf k}_{1}^{2})F_{\beta}({\bf k}_{2}^{2}) \nonumber\\
  && \times \sum_{p}O^{(n)}_{\alpha\beta,p}({\bf k}_{1},\omega_{1};
        {\bf k}_{2},\omega_{2})D_{p}^{(n)}(\omega_{1},\omega_{2}),
                                                          \nonumber\\
  && \label{Vpairmom}
\end{eqnarray}
where the index $n$ distinguishes one-pair ($n=1$) and two-pair
($n=2$) meson-pair exchange, and ($\alpha\beta$) refers to the particular
meson pair that is being exchanged. The product of the coupling constants
in the two cases is given by
\begin{eqnarray}
  g^{(1)}(\alpha\beta) &=& g_{(\alpha\beta)}
                           g_{N\!N\alpha}g_{N\!N\beta}, \nonumber\\
  g^{(2)}(\alpha\beta) &=& g^{2}_{(\alpha\beta)},
\end{eqnarray}
with appropriate powers of $m_{\pi}$, depending on the definition
of the Hamiltonians. The energy denominators $D^{(n)}_{p}$ are
given in Eq.~(\ref{Dpairs}), with the index $p$ labeling the
different time-ordered processes. Finally, the momentum-dependent
operators $O^{(n)}_{\alpha\beta,p}$ are given in Tables~\ref{O1pair} and
\ref{O2pair}. For completeness, these Tables also contain the isospin
factors $C^{(n)}(\alpha\beta)$ as derived in Appendix~\ref{app:appB}.
The momentum operators for $(\pi\pi)_{0}$ and $(\pi\pi)_{1}$ both
contain a term antisymmetric in ${\bf k}_{1}\leftrightarrow{\bf k}_{2}$.
They only contribute when we make the nonadiabatic expansion (see
Sec.~\ref{sec:chap4}). In the leading-order potential (\ref{Vpairmom})
they drop out when we integrate over ${\bf k}_{1}$ and ${\bf k}_{2}$.

The time-ordering label $p$ in Eq.~(\ref{Vpairmom}) is only of interest
for the $(\pi\pi)_{1}$ one-pair potential, where we have an explicit
$(\omega_{1},\omega_{2})$ dependence in $O^{(1)}_{\pi\pi,p}$.
For all other one-pair potentials we have the same energy denominator
$D^{(1)}(\omega_{1},\omega_{2})$ as given in Eq.~(\ref{D1pair}), and we
can drop the $p$ index in $O^{(1)}_{\alpha\beta,p}$.
For $(\alpha\beta)=(\pi\pi)_{1}$, the three time-ordered diagrams
contribute as
\begin{eqnarray*}
  &&  (-\omega_{1}+\omega_{2})D^{(1)}_{a}+
       (\omega_{1}+\omega_{2})D^{(1)}_{b}+
       (\omega_{1}-\omega_{2})D^{(1)}_{c}            \nonumber\\
  &&  \hspace*{1cm}  = \frac{2}{\omega_{1}\omega_{2}
                               (\omega_{1}+\omega_{2})}.
\end{eqnarray*}
Similarly, the two-pair potentials are all seen to have the energy
denominator $D^{(2)}$ of Eq.~(\ref{Dpairs}), where the additional
$(\omega_{1}-\omega_{2})^{2}$ dependence in $O^{(2)}_{\alpha\beta}$ for
$(\pi\pi)_{1}$ can be rewritten as
\[
  (\omega_{1}-\omega_{2})^{2}D^{(2)} =-\frac{1}{2\omega_{1}}
        -\frac{1}{2\omega_{2}}
        +\frac{2}{\omega_{1}+\omega_{2}}.
\]

The evaluation of the momentum integrations can now readily be
performed using the methods given in~\cite{Rij95a,Rij91}.
There it was shown that the full separation of the ${\bf k}_{1}$
and ${\bf k}_{2}$ dependence of the Fourier integrals can be
achieved in all cases using the $\lambda$-integral representation.
Starting-out from Eq.~(\ref{Vpairmom}), this procedure gives the
following generic form of the potentials in coordinate space:
\begin{eqnarray}
   V^{(n)}_{\rm pair}(\alpha\beta) &=&
         C^{(n)}(\alpha\beta) g^{(n)}(\alpha\beta)
         \lim_{{\bf r}_{1},{\bf r}_{2}\rightarrow{\bf r}}   \nonumber\\
   &&\times O^{(n)}_{\alpha\beta}(-i\bbox{\nabla}_{1},\omega_{1};
       -i\bbox{\nabla}_{2},\omega_{2})
       B^{(n)}_{\alpha\beta}(r_{1},r_{2}).  \nonumber\\
   & & \label{Vpaircor}
\end{eqnarray}
The different functions $B_{\alpha\beta}(r_{1},r_{2})$ that occur in
this expression involve the functions $I_{2}(m,r)$ as defined in
Ref.~\cite{Rij91}, $I_{0}(\Lambda,r)$ as defined in Ref.~\cite{Rij95a},
and the integrals
\begin{eqnarray}
   B_{0,0}(m_{\alpha},r_{1};m_{\beta},r_{2}) &\!=\!& \frac{2}{\pi}
      \int_{0}^{\infty}\!d\lambda\,\lambda^{2}F_{\alpha}(\lambda,r_{1})
               F_{\beta}(\lambda,r_{2}), \nonumber\\
   B_{1,1}(m_{\alpha},r_{1};m_{\beta},r_{2}) &\!=\!& \frac{2}{\pi}
      \int_{0}^{\infty}\!d\lambda F_{\alpha}(\lambda,r_{1})
               F_{\beta}(\lambda,r_{2}),
\end{eqnarray}
where
\begin{equation}
   F_{\alpha}(\lambda,r)=e^{-\lambda^{2}/\Lambda_{\alpha}^{2}}
           I_{2}(\sqrt{m_{\alpha}^{2}+\lambda^{2}},r).
\end{equation}
The differentiation operations needed in Eq.~(\ref{Vpaircor}) are
listed in Appendix A of the previous paper~\cite{Rij95a}, and the
resulting coordinate-space potentials are here given explicitly in
Appendix~\ref{app:appC}.

\section{1/M CORRECTIONS}
\label{sec:chap4}
The nonadiabatic correction from the $1/M$ expansion of the energy
denominators is explained in Ref.~\cite{Rij91}. The expansion of
the energy denominator involves a momentum dependence which can be
rewritten in the form $[{\bf k}_{1}\!\cdot\!{\bf k}_{2}-
{\bf q}\!\cdot\!({\bf k}_{1}-{\bf k}_{2})]/2M$.

Taking out the momentum-dependent factor, the one-pair energy
denominators of Eq.~(\ref{Dpairs}) give rise to the energy denominators
\begin{eqnarray}
 D^{\rm na}_{a}(\omega_{1},\omega_{2})&=&\frac{1}{2\omega_{1}\omega_{2}}
     \,\frac{1}{\omega^{2}_{2}(\omega_{1}+\omega_{2})},   \nonumber\\
 D^{\rm na}_{b}(\omega_{1},\omega_{2})&=&\frac{1}{2\omega_{1}\omega_{2}}
       \left(\frac{1}{\omega_{1}^{2}\omega_{2}}
            +\frac{1}{\omega_{1}\omega_{2}^{2}}\right),    \nonumber\\
 D^{\rm na}_{c}(\omega_{1},\omega_{2})&=&\frac{1}{2\omega_{1}\omega_{2}}
     \,\frac{1}{\omega^{2}_{1}(\omega_{1}+\omega_{2})}.  \label{Dnadia}
\end{eqnarray}
Again, the time ordering is only of importance in the $(\pi\pi)_{1}$
potential, where the contributions sum up to
\begin{equation}
  D^{\rm na}_{(\pi\pi)_{1}} = \frac{2}{\omega_{1}^{2}\omega_{2}^{2}}.
\end{equation}
In all other cases the three energy denominators of Eq.~(\ref{Dnadia})
can be summed directly, yielding
\begin{equation}
  D^{\rm na} = \frac{1}{\omega_{1}^{2}\omega_{2}^{2}}
       \left[\frac{1}{\omega_{1}}+\frac{1}{\omega_{2}}
            -\frac{1}{\omega_{1}+\omega_{2}}\right],
\end{equation}
which gives rise to a coordinate-space function of the form
\begin{eqnarray}
  B^{\rm na}_{\alpha\beta}(r_{1},r_{2}) = \frac{2}{\pi}
      \int_{0}^{\infty}\!\frac{d\lambda}{\lambda^{2}}  &&
      \biggl[I_{2}(m_{\alpha},r_{1})I_{2}(m_{\beta},r_{2})  \nonumber\\
  &&     -F_{\alpha}(\lambda,r_{1})F_{\beta}(\lambda,r_{2})\biggr].
                                                           \label{Bna}
\end{eqnarray}

The energy denominators $D^{\rm na}(\omega_{1},\omega_{2})$ are
symmetric under interchanges of labels 1 and 2, and so the term
proportional to ${\bf q}\!\cdot\!({\bf k}_{1}-{\bf k}_{2})$ will
only contribute in combination with the antisymmetric terms in
$O^{(1)}({\bf k}_{1},\omega_{1};{\bf k}_{2}\omega_{2})$; i.e., in
$(\pi\pi)_{0}$ and $(\pi\pi)_{1}$. We find
\widetext
\begin{eqnarray}
   V^{\rm na}[(\pi\pi)_{0}] &=& -\frac{g_{(\pi\pi)_{0}}}{m_{\pi}}
       \left(\frac{f_{N\!N\pi}}{m_{\pi}}\right)^{2}\frac{3}{M}
     \int\!\!\int\!\frac{d^{3}k_{1}d^{3}k_{2}}{(2\pi)^{6}} \,
     e^{i({\bf k}_{1}+{\bf k}_{2})\cdot{\bf r}}
     F_{\pi}({\bf k}_{1}^{2})F_{\pi}({\bf k}_{2}^{2})      \nonumber\\
   & & \times \biggl[ ({\bf k}_{1}\!\cdot\!{\bf k}_{2})^{2}+
       {\textstyle\frac{i}{2}}{\bf q}\!\cdot\!({\bf k}_{1}-{\bf k}_{2})
       (\bbox{\sigma}_{1}+\bbox{\sigma}_{2})\!\cdot\!
       ({\bf k}_{1}\times{\bf k}_{2}) \biggr]
       D^{\rm na}(\omega_{1},\omega_{2}),           \label{Vnapipi0}\\
   V^{\rm na}[(\pi\pi)_{1}] &=&-(\bbox{\tau}_{1}\!\cdot\!\bbox{\tau}_{2})
       \frac{g_{(\pi\pi)_{1}}}{m_{\pi}^{2}}
       \left(\frac{f_{N\!N\pi}}{m_{\pi}}\right)^{2}\frac{1}{M}
     \int\!\!\int\!\frac{d^{3}k_{1}d^{3}k_{2}}{(2\pi)^{6}} \,
     e^{i({\bf k}_{1}+{\bf k}_{2})\cdot{\bf r}}
     F_{\pi}({\bf k}_{1}^{2})F_{\pi}({\bf k}_{2}^{2})      \nonumber\\
   & & \times \biggl[ ({\bf k}_{1}\!\cdot\!{\bf k}_{2})^{2}+
       {\textstyle\frac{i}{2}}{\bf q}\!\cdot\!({\bf k}_{1}-{\bf k}_{2})
       (\bbox{\sigma}_{1}+\bbox{\sigma}_{2})\!\cdot\!
       ({\bf k}_{1}\times{\bf k}_{2}) \biggr]
       \frac{2}{\omega_{1}^{2}\omega_{2}^{2}},      \label{Vnapipi1}\\
   V^{\rm na}[(\sigma\sigma)] &=& \frac{g_{(\sigma\sigma)}}{m_{\pi}}\,
       \frac{g^{2}_{N\!N\sigma}}{M}
     \int\!\!\int\!\frac{d^{3}k_{1}d^{3}k_{2}}{(2\pi)^{6}} \,
     e^{i({\bf k}_{1}+{\bf k}_{2})\cdot{\bf r}}
     F_{\sigma}({\bf k}_{1}^{2})F_{\sigma}({\bf k}_{2}^{2})
     ({\bf k}_{1}\!\cdot\!{\bf k}_{2})
     D^{\rm na}(\omega_{1},\omega_{2}),             \label{Vnasisi}\\
   V^{\rm na}[(\pi\sigma)] &=&-(\bbox{\tau}_{1}\!\cdot\!\bbox{\tau}_{2})
      \frac{g_{(\pi\sigma)}}{m_{\pi}^{2}}\,
      \frac{f_{N\!N\pi}}{m_{\pi}}\, \frac{g_{N\!N\sigma}}{M}
      \int\!\!\int\!\frac{d^{3}k_{1}d^{3}k_{2}}{(2\pi)^{6}}\,
      e^{i({\bf k}_{1}+{\bf k}_{2})\cdot{\bf r}}
      F_{\pi}({\bf k}_{1}^{2})F_{\sigma}({\bf k}_{2}^{2}) \nonumber\\
   & & \times ({\bf k}_{1}\!\cdot\!{\bf k}_{2}) \biggl[
           \bbox{\sigma}_{1}\!\cdot\!{\bf k}_{1}
           \bbox{\sigma}_{2}\!\cdot\!{\bf k}_{1}-{\textstyle\frac{1}{2}}
          (\bbox{\sigma}_{1}\!\cdot\!{\bf k}_{1}
           \bbox{\sigma}_{2}\!\cdot\!{\bf k}_{2}
          +\bbox{\sigma}_{1}\!\cdot\!{\bf k}_{2}
           \bbox{\sigma}_{2}\!\cdot\!{\bf k}_{1}) \biggr]
           D^{\rm na}(\omega_{1},\omega_{2}),         \label{Vnapisi}
\end{eqnarray}
where we substituted the isospin dependence and the coupling constants.

The pseudovector vertex gives rise to $1/M$ terms as shown in
Eq.~(\ref{Gam1p}). Taking into account the different time orderings
and discarding the terms antisymmetric under interchange of
${\bf k}_{1}\leftrightarrow{\bf k}_{2}$, we find the following
contributions
\begin{eqnarray}
  V^{\rm pv}[(\pi\pi)_{0}] &=& \frac{g_{(\pi\pi)_{0}}}{m_{\pi}}
     \left(\frac{f_{N\!N\pi}}{m_{\pi}}\right)^{2}\frac{3}{M}
     \int\!\!\int\!\frac{d^{3}k_{1}d^{3}k_{2}}{(2\pi)^{6}} \,
     e^{i({\bf k}_{1}+{\bf k}_{2})\cdot{\bf r}}
     F_{\pi}({\bf k}_{1}^{2})F_{\pi}({\bf k}_{2}^{2})      \nonumber\\
  & & \times \biggl[ {\bf k}_{1}^{2}+{\bf k}_{2}^{2}-
          i(\bbox{\sigma}_{1}+\bbox{\sigma}_{2})\!\cdot\!
          ({\bf k}_{1}+{\bf k}_{2})\times{\bf q} \biggr] \frac{1}
       {\omega_{1}\omega_{2}(\omega_{1}+\omega_{2})}, \label{Vpvpipi0}\\
  V^{\rm pv}[(\pi\pi)_{1}] &=& (\bbox{\tau}_{1}\!\cdot\!\bbox{\tau}_{2})
      \frac{g_{(\pi\pi)_{1}}}{m_{\pi}^{2}}
      \left(\frac{f_{N\!N\pi}}{m_{\pi}}\right)^{2}\frac{1}{M}
      \int\!\!\int\!\frac{d^{3}k_{1}d^{3}k_{2}}{(2\pi)^{6}}\,
      e^{i({\bf k}_{1}+{\bf k}_{2})\cdot{\bf r}}
      F_{\pi}({\bf k}_{1}^{2})F_{\pi}({\bf k}_{2}^{2}) \nonumber\\
  && \times \left[\left(\frac{{\bf k}_{1}^{2}}{\omega_{1}^{2}}+
                  \frac{{\bf k}_{2}^{2}}{\omega_{2}^{2}}\right)
    -i(\bbox{\sigma}_{1}+\bbox{\sigma}_{2})\!\cdot\!\left(
     \frac{{\bf k}_{1}}{\omega_{1}^{2}}+
                  \frac{{\bf k}_{2}}{\omega_{2}^{2}}\right)
            \!\times\!{\bf q}\right],               \label{Vpvpipi1}\\
  V^{\rm pv}[(\pi\sigma)] &=&-(\bbox{\tau}_{1}\!\cdot\!\bbox{\tau}_{2})
      \frac{g_{(\pi\sigma)}}{m_{\pi}^{2}}\,
      \frac{f_{N\!N\pi}}{m_{\pi}}\, \frac{g_{N\!N\sigma}}{M}
   \int\!\!\int\!\frac{d^{3}k_{1}d^{3}k_{2}}{(2\pi)^{6}}\,
    e^{i({\bf k}_{1}+{\bf k}_{2})\cdot{\bf r}}
      F_{\pi}({\bf k}_{1}^{2})F_{\sigma}({\bf k}_{2}^{2}) \nonumber\\
   & & \times \biggl[ \bbox{\sigma}_{1}\!\cdot\!{\bf k}_{2}
           \bbox{\sigma}_{2}\!\cdot\!{\bf k}_{2}-{\textstyle\frac{1}{2}}
          (\bbox{\sigma}_{1}\!\cdot\!{\bf k}_{1}
           \bbox{\sigma}_{2}\!\cdot\!{\bf k}_{2}
          +\bbox{\sigma}_{1}\!\cdot\!{\bf k}_{2}
           \bbox{\sigma}_{2}\!\cdot\!{\bf k}_{1}) \biggr]
      \frac{1}{\omega_{1}\omega_{2}(\omega_{1}+\omega_{2})}.
                                                      \label{Vpvpisi}
\end{eqnarray}
\narrowtext\noindent

\section{THEORETICAL CONSTRAINTS\protect\\
         ON PAIR COUPLING CONSTANTS}
\label{sec:chap5}
In principle, each of the pair-meson potentials contains a free
parameter, the pair-meson coupling constant. A possible way to fix
the values for the pair-meson coupling constants is to assume that the
coupling of a meson pair $(\alpha\beta)$ to a nucleon is dominated by
an intermediate boson $H$. We can then relate the pair coupling
constant $g_{(\alpha\beta)}$ to the meson-decay and meson-nucleon
coupling constants, $g_{H\alpha\beta}$ and $g_{N\!NH}$, respectively.
The relevant nucleon-nucleon-meson and meson-decay Lagrangians can
then be obtained by extending the linear $\sigma$ model~\cite{Gel60}
with vector and axial-vector mesons; see, for example,
Refs.~\cite{Gas69,Lee72,Ko94}.

Meson saturation is graphically presented in Fig.~\ref{pap2fig4}.
Because of the additional meson propagator and the meson-decay vertex,
we find the relation
\begin{equation}
   g_{(\alpha\beta)} \approx -\frac{(-)^{S}}{m_{H}^{2}}
                      g_{N\!N\!H}g_{H\alpha\beta},   \label{gsatur}
\end{equation}
with appropriate powers of meson masses, depending on the actual form
of the meson-pair and meson-decay Hamiltonians. Here $S$ denotes the
spin of the intermediate meson $H$ and we assumed that $m_{H}^{2}$
is much larger than the momentum-transfer squared. However, for the
$(\pi\pi)_{1}$ and $(\pi\pi)_{0}$ pairs the decay can proceed via
the $\rho$ and $\sigma$ mesons, respectively. These are substantially
lighter than 1 GeV, and so we are faced with the difficult task of
finding a suitable average momentum squared to approximate
$({\bf k}^{2}+m_{H}^{2})$.
Another problem is that a specific meson pair might be the decay
product of a range of different mesons, and so Eq.~(\ref{gsatur})
becomes a sum over mesons $H_{i}$, confusing the issue even further.
Alternatively, we can also use the linear $\sigma$ model to generate
the pair vertices explicitly, which we shall do in the following.

The $\sigma$ model has been discussed extensively in the literature,
and here we only briefly outline its contents to define the quantities
we need. The model contains an isotriplet of pseudoscalars, $\bbox{\pi}$,
and an isosinglet scalar, $\sigma$, grouped into
\begin{equation}
   \Sigma=\sigma-i\bbox{\tau}\!\cdot\!\bbox{\pi},
\end{equation}
which transforms under global SU(2)$_{L}$$\times$SU(2)$_{R}$ as
\begin{equation}
   \Sigma \rightarrow  L\Sigma R^{\dagger}.
\end{equation}
The nucleon wave function $\psi$ has left and right components,
$\psi_{L,R}={\textstyle\frac{1}{2}}(1\mp\gamma_{5})\psi$,
transforming as
\begin{equation}
  \psi_{L}\rightarrow L\psi_{L}, \ \ \ \psi_{R}\rightarrow R\psi_{R}.
\end{equation}
The transformations can be made local by introducing the left and
right gauge fields,
\begin{eqnarray}
  l_{\mu}&\equiv&{\textstyle\frac{1}{2}}\bbox{\tau}\!\cdot\!{\bf l}_{\mu}
       ={\textstyle\frac{1}{2}}\bbox{\tau}\!\cdot\!
        (\bbox{\rho}_{\mu}-{\bf a}_{\mu}),      \nonumber\\
  r_{\mu}&\equiv&{\textstyle\frac{1}{2}}\bbox{\tau}\!\cdot\!{\bf r}_{\mu}
       ={\textstyle\frac{1}{2}}\bbox{\tau}\!\cdot\!
        (\bbox{\rho}_{\mu}+{\bf a}_{\mu}),
\end{eqnarray}
and their field strength tensors
\begin{equation}
  l_{\mu\nu}=\partial_{\mu}l_{\nu}-\partial_{\nu}l_{\mu}
            +ig_{V}[l_{\mu},l_{\nu}],
\end{equation}
and similarly for $r_{\mu\nu}$. The vector and axial vector mesons are
given a mass by introducing a chiral-symmetry breaking mass term
\begin{equation}
   {\cal L}_{m}={\textstyle\frac{1}{2}}
         m_{\rho}^{2}{\rm Tr}(l_{\mu}l^{\mu}+r_{\mu}r^{\mu}).
\end{equation}
After the introduction of the gauge fields, the chiral symmetry can be
restored by defining the following covariant derivatives for the
nucleon and $(\sigma,\bbox{\pi})$ fields,
\begin{eqnarray}
 {\cal D}_{\mu}\psi &=& (\partial_{\mu}+
     {\textstyle\frac{i}{2}}g_{V}\bbox{\tau}\!\cdot\!\bbox{\rho}_{\mu}
    +{\textstyle\frac{i}{2}}g_{V}\bbox{\tau}\!\cdot\!{\bf a}_{\mu}
      \gamma_{5})\psi,                                      \nonumber\\
 {\cal D}_{\mu}\Sigma   &=& \partial_{\mu}\Sigma
       +ig_{V}l_{\mu}\Sigma-ig_{V}\Sigma r_{\mu}.
\end{eqnarray}

A possible chiral-invariant Lagrangian is now given by
\begin{eqnarray}
 {\cal L}_{0} &=& \bar{\psi}i\gamma^{\mu}{\cal D}_{\mu}\psi
            -g_{1}(\bar{\psi}_{L}\Sigma\psi_{R}
                  +\bar{\psi}_{R}\Sigma^{\dagger}\psi_{L})  \nonumber\\
 && +g_{2}(\bar{\psi}_{L}\Sigma\Sigma^{\dagger}\Sigma\psi_{R}
          +\bar{\psi}_{R}\Sigma^{\dagger}\Sigma\Sigma^{\dagger}\psi_{L})
                                                            \nonumber\\
 && +iC(\bar{\psi}_{L}\Sigma\gamma^{\mu}{\cal D}_{\mu}
                     \Sigma^{\dagger}\psi_{L}
       +\bar{\psi}_{R}\Sigma^{\dagger}\gamma^{\mu}{\cal D}_{\mu}
                     \Sigma\psi_{R})                         \nonumber\\
 && -{\textstyle\frac{1}{4}}{\rm Tr}(l_{\mu\nu}l^{\mu\nu}
                                    +r_{\mu\nu}r^{\mu\nu})
    +{\textstyle\frac{1}{4}}{\rm Tr}({\cal D}^{\mu}\Sigma^{\dagger}
               {\cal D}_{\mu}\Sigma)                         \nonumber\\
 && -{\textstyle\frac{1}{4}}\mu^{2}{\rm Tr}(\Sigma^{\dagger}\Sigma)
    -{\textstyle\frac{1}{8}}\lambda^{2}{\rm Tr}
     (\Sigma^{\dagger}\Sigma)^{2},                   \label{Lagchiralb}
\end{eqnarray}
where $g_{1}$, $g_{2}$, $C$, $\mu$, and $\lambda$ are free parameters.

The symmetry can be spontaneously broken by adding a term linear in
the $\sigma$ field,
\begin{equation}
     {\cal L}_{SB}=f_{\pi}m_{\pi}^{2}\sigma,
\end{equation}
with $f_{\pi}=92.4$ MeV the pion decay constant.
Choosing the ground state as $\langle\sigma\rangle=v$, this spontaneous
symmetry breaking introduces a mixing between the $\bbox{\pi}$ and
${\bf a}_{\mu}$ fields, which requires a redefinition of the fields,
given by
\begin{eqnarray}
   \sigma &\rightarrow& \sigma+v, \nonumber\\
   {\bf a}_{\mu} &\rightarrow& {\bf A}_{\mu}-\frac{g_{V}v}{m_{A}^{2}}
                 D_{\mu}\bbox{\pi}, \nonumber\\
   \bbox{\pi} &\rightarrow& \frac{m_{A}}{m_{\rho}}\bbox{\pi},
\end{eqnarray}
where we defined $m_{A}^{2}=m_{\rho}^{2}+(g_{V}v)^{2}$,
$v=(m_{A}/m_{\rho})f_{\pi}$, and $D_{\mu}\bbox{\pi}=\partial_{\mu}
\bbox{\pi}+g_{V}\bbox{\pi}\times\bbox{\rho}_{\mu}$.
The shift of the $\sigma$ field gives the nucleons a mass
$M=(g_{1}-g_{2}v^{2})v$, which imposes a constraint on $g_{1}$ and
$g_{2}$.

Comparing the various pieces of the full Lagrangian, ${\cal L}_{0}
+{\cal L}_{m}+{\cal L}_{SB}$, with the interaction Hamiltonians of
Sec.~\ref{sec:chap2}, we identify the following relations for the
single-meson coupling constants:
\begin{eqnarray}
   g_{N\!N\sigma} &=& (g_{1}-3g_{2}v^{2})=M/v-2g_{2}v^{2}, \nonumber\\
   f_{N\!N\pi} &=& \frac{m_{\pi}}{2f_{\pi}}
     \left(2Cv^{2}-\frac{m_{\rho}^{2}}{m_{A}^{2}}\right), \nonumber\\
   g_{N\!N\rho} &=& {\textstyle\frac{1}{2}}g_{V},   \nonumber\\
   g_{N\!NA} &=& \frac{m_{A}}{m_{\pi}}
             \frac{\sqrt{m_{A}^{2}-m_{\rho}^{2}}}{m_{\rho}}f_{N\!N\pi}.
\end{eqnarray}
The numerical value for $C$ is obtained by imposing the
Goldberger-Treiman~\cite{Gol58} relation,
\begin{equation}
  \frac{f_{N\!N\pi}}{m_{\pi}}=\frac{g_{A}}{2f_{\pi}},
\end{equation}
with $g_{A}=1.2573$ the weak interaction axial-vector coupling
constant. Similarly, the meson-pair coupling constants are found to be
\begin{eqnarray}
  g_{(\pi\pi)_{0}} &=& -\frac{m_{\pi}}{2f_{\pi}}\left(
    \frac{M}{f_{\pi}}-g_{N\!N\sigma}\frac{m_{A}}{m_{\rho}}
    \right),                                             \nonumber\\
  g_{(\sigma\sigma)} &=& 3g_{(\pi\pi)_{0}}\,
                   \frac{m_{\rho}^{2}}{m_{A}^{2}},       \nonumber\\
  g_{(\pi\pi)_{1}} &=& \frac{m_{\pi}^{2}}{2f_{\pi}^{2}}\,
                   \frac{m_{\rho}^{2}}{m_{A}^{2}}
     \left(g_{A}+\frac{m_{\rho}^{2}}{m_{A}^{2}}\right),  \nonumber\\
  g_{(\pi\rho)_{1}} &=& \frac{g_{N\!N\rho}}{2f_{\pi}}
                      (g_{A}+1),                         \nonumber\\
  g_{(\pi\sigma)} &=& \frac{m_{\pi}^{2}}{2f_{\pi}^{2}}\,
           \frac{m_{A}^{2}-2m_{\rho}^{2}}{m_{A}m_{\rho}}
           \left(g_{A}+\frac{m_{\rho}^{2}}{m_{A}^{2}}\right).
\end{eqnarray}

Working-out the various pieces of the Lagrangian we can also find
relations for the meson-decay coupling constants. It is then found that
the relation (\ref{gsatur}) is indeed satisfied, although in some cases
the approximation is rather crude, as we already suspected.
However, the meson-saturation assumption is still very useful to
suggest the obvious relation
\begin{equation}
   f_{(\pi\pi)_{1}} = g_{(\pi\pi)_{1}}\,(f_{N\!N\rho}/g_{N\!N\rho}).
\end{equation}

\section{RESULTS AND DISCUSSION}
\label{sec:chap6}
The complete one-pair and two-pair potential can be written as
\begin{eqnarray}
  V(\alpha\beta)&=&V^{(1)}(\alpha\beta)+V^{(2)}(\alpha\beta)\nonumber\\
     &&+V^{\rm na}(\alpha\beta)+V^{\rm pv}(\alpha\beta),
\end{eqnarray}
for the pairs $(\alpha\beta)$ discussed in this paper.
To present the potentials in graphical form, we employ the meson-nucleon
coupling constants and cutoff masses of a preliminary version of the
Nijmegen extended soft-core (ESC) potential. The ESC potential is still
under construction, but the present values for the coupling constants
clearly suffice for illustrative purposes.
The meson-nucleon coupling constants and cutoff masses are already
listed in Table IV of the previous paper~\cite{Rij95a}, while the pair
coupling constants are here given in Table~\ref{gpair}. All pair
coupling constants are fixed at their theoretical values as given in
Sec.~\ref{sec:chap5}. This constraint might be a bit too severe, but
the values can at least be expected to be reasonable. Furthermore,
in this way we can get a feeling about the importance of two-meson
exchanges with respect to one-boson exchanges, because we do not
introduce any new free parameters.

In Figs.~\ref{pap2fig5} and \ref{pap2fig6}, we compare the various
types of one-pair and two-pair exchanges for $I=0$ and $I=1$,
respectively.
They consist of scalar $0^{++}$ exchanges [$(\pi\pi)_{0}$ and
$(\sigma\sigma)$], vector $1^{--}$ exchange [$(\pi\pi)_{1}$], and
axial $1^{++}$ exchanges [$(\pi\rho)_{1}$ and $(\pi\sigma)$].
The central $I=0$ potential is clearly dominated by the one-pair
$(\pi\pi)_{1}$ potential, whereas the scalar and vector two-pair
potentials largely cancel each other. For $I=1$, the scalar one-pair
and two-pair potentials are the same, whereas the vector potentials
change by a factor --3, resulting in a much smaller (but still
attractive) central potential. For the spin-spin and tensor
potentials the $I=0$ and $I=1$ cases differ by a simple factor
of --3, where the two-pair axial tensor component turns out to
be completely negligible.

In all cases, the $1/M$ corrections from the nonadiabatic expansion
of the energy denominators and from the pseudovector-vertex correction
in the pion vertex function have the opposite sign. In the
$(\pi\pi)_{0}$ potential the cancelation is even exact.
The $1/M$ corrections to the spin-spin and tensor potentials are only
due to the $(\pi\sigma)$ potential, whereas the $(\pi\pi)_{0}$ and
$(\pi\pi)_{1}$ potentials only contribute to the central and spin-orbit
potentials.

In Fig.~\ref{pap2fig7} we compare the spin-orbit components from the
one-pair potentials. They are all due to the $1/M$ nonadiabatic or
pseudovector-vertex corrections in the $(\pi\pi)_{0}$ and $(\pi\pi)_{1}$
potentials. The $(\pi\pi)_{1}$ potential also has a spin-orbit potential
due to the $1/M$ term in the $(\pi\pi)_{1}$ pair vertex (\ref{Gam2v}),
as given in Eq.~(\ref{tmep7}) of Appendix~\ref{app:appC}.
This contribution is exactly the same as the nonadiabatic contribution,
which means that in the $(\pi\pi)_{1}$ potential there is an even larger
cancelation between the nonadiabatic (long-dashed line) and
pseudovector-vertex (dash-dotted line) spin-orbit parts than
Fig.~\ref{pap2fig7} might suggest. Note that in the $(\pi\pi)_{0}$
potential the cancelation between the nonadiabatic (solid line) and
pseudovector-vertex (short-dashed line) spin-orbit parts is exact,
and so there is no spin-orbit potential from $0^{++}$ pair exchange.
Obviously, if we were to take the pseudoscalar coupling,
$\bar{\psi}i\gamma_{5}\psi\phi$, for the pion, rather than the
pseudovector coupling, $\bar{\psi}\gamma_{5}\gamma_{\mu}\psi
\partial^{\mu}\phi$, these cancelations do not occur, and we are left
with the spin-orbit contribution from the nonadiabatic expansion.

Finally, we should mention that in some cases the two-pair diagrams
are in principle included in the cases where we use the exchange of
broad mesons. This is the case for the two-pair $(\pi\pi)_{0}$ and
$(\pi\pi)_{1}$ potentials. Here we use in the Nijmegen work a broad
$\varepsilon$ and $\rho$ one-boson-exchange potential. If such a broad
meson exchange is included exactly, the two-pair $\pi\pi$ potentials
should be omitted. When such broad mesons are included in a two-pole
approximation~\cite{Sch71,Bin71}, however, a two-pair contribution
might still be useful. Whether these two-pair contributions in that
case should be included in full or partially suppressed is presently
under investigation.

Let us now return to a comparison of this new ESC potential with
the one-boson-exchange Nijm93 potential~\cite{Sto94}. It is important
to realize that both models contain essentially the same set of
parameters, and so the ESC model is a true extension of the Nijm93
model. Using the linear $\sigma$ model as a means to get a reasonable
estimate for the various meson-pair coupling constants, however, the
contributions from the two-meson exchange diagrams (the planar and
crossed-box diagrams as evaluated in the previous paper~\cite{Rij95a}
as well as the one-pair and two-pair diagrams as evaluated in the
present paper) are included without the introduction of any new
parameters. The 14 free parameters of the ESC model are fitted to
the 1993 Nijmegen representation of the $\chi^{2}$ hypersurface of
the $N\!N$ scattering data below $T_{\rm lab}=350$ MeV~\cite{Sto93},
updated with the inclusion of new data which have been published
since then.

The results for the 10 energy bins are given in Table~\ref{tabchi2},
where we compare the results from the updated partial-wave analysis
with the Nijm93 and ESC potentials. Clearly, the inclusion of the
two-meson exchanges in the ESC model allows for a substantially
better description of the $N\!N$ scattering data than what could
be achieved with the one-boson-exchange Nijm93 model. At present,
the ESC potential is the only meson-theoretical model which can
give such a good description of the scattering data (using only
a limited set of free parameters). The quality of the ESC potential
is even better if we restrict its application to energies below
$T_{\rm lab}\approx300$ MeV, giving $\chi^{2}/N_{\rm data}=1.137$
for the 0--290 MeV energy interval. This was to be expected, since
the nonadiabatic expansion in principle is only valid below the
pion-production threshold.

To summarize, the combined results of the present and the
previous~\cite{Rij95a} paper show that it is possible to construct a
nucleon-nucleon potential model, based on a chiral-symmetric Lagrangian,
that gives a good description of the nucleon-nucleon scattering data.
Fine-tuning of the 14 parameters of this extended model allows for a
substantially better description of the data than what was possible
with the 14-parameter one-boson-exchange Nijm93 model~\cite{Sto94}.

\acknowledgments
We would like to thank Prof.\ J.J.\ de Swart, Prof.\ I.R.\ Afnan and
the other members of the theory groups at Nijmegen and Flinders for
their stimulating interest.
The work of V.S.\ was financially supported by the Australian Research
Council.

\widetext
\appendix
\section{ENERGY INTEGRALS}
\label{app:appA}
We discuss here the treatment of the energy integrals that occur
in the evaluation of the meson-pair potentials.
To prevent displaying all the indices, we introduce the following
convenient notations
\begin{eqnarray}
 &&  \omega=\sqrt{{\bf k}^{2}+m^{2}},          \ \ \
     \omega'=\sqrt{{\bf k}'^{2}+m'^{2}},      %\ \ \
\nonumber\\ &&
     A=E({\bf p})-{\textstyle\frac{1}{2}} W,   \ \ \
     A'=E({\bf p}')-{\textstyle\frac{1}{2}} W, \ \ \
     A''=E({\bf p}'')-{\textstyle\frac{1}{2}} W,     \label{Adefpar}
\end{eqnarray}
where ${\bf p}'-{\bf p}={\bf k}+{\bf k}'$ and
${\bf p}''={\bf p}+{\bf k}$. Note that with this notation, ${\bf k}$
and ${\bf k}'$ correspond to ${\bf k}_{1}$ and ${\bf k}_{2}$,
respectively, introduced in the main text.

(i) {\it The one-pair graph.}
Here we encounter the integral
\begin{eqnarray}
  {\cal J}_{1P}({\bf p}',{\bf p}|W) &=& -(2\pi)^{-2}
     \left[ W-2E({\bf p}')\right]\left[W-2E({\bf p})\right]
\nonumber\\ && \times
     \int\!dp'_{0}\int\!dp_{0}\int\!dk'_{0}\int\!dk_{0}\
        \delta^{4}(p'-p-k-k')                           \nonumber\\
  && \times [k^{\prime 2}-m^{\prime 2}+i\delta]^{-1}\,
           \left[F^{(a)}_{W}({\bf p}',p'_{0})^{-1}\
          F^{(b)}_{W}(-{\bf p}',-p'_{0})\right]^{-1}    \nonumber\\
  && \times \left[F^{(b)}_{W}(-{\bf p}-{\bf k},
                            -p_{0}-k_{0})\right]^{-1}   \nonumber\\
  && \times [k^{2}-m^{2}+i\delta]^{-1}\,
           \left[F^{(a)}_{W}({\bf p},p_{0})\
                F^{(b)}_{W}(-{\bf p},-p_{0})\right]^{-1}.
\end{eqnarray}
Similar integrals were treated in \cite{Rij91}, to which we refer
for the method of evaluation. We first bring the integral into the form
\begin{eqnarray}
  {\cal J}_{1P} = (2\pi)^{-4} [4A'A]\ &&
     \int_{-\infty}^{+\infty}\!\!d\alpha
     \int_{-\infty}^{+\infty}\!\!d\beta
     \int_{-\infty}^{+\infty}\!\!dp'_{0}
     \int_{-\infty}^{+\infty}\!\!dp_{0}
     \int_{-\infty}^{+\infty}\!\!dk'_{0}
     \int_{-\infty}^{+\infty}\!\!dk_{0}
     \int_{-\infty}^{+\infty}\!\!dp''_{0}\
\nonumber\\ && \times
    e^{i\alpha(p'_{0}-k'_{0}-p''_{0})}
    e^{i\beta(p''_{0}-p_{0}-k_{0})}                        \nonumber\\
  && \times [\omega'^{2}-k_{0}^{\prime 2}-i\delta]^{-1}
            [\omega^{2}-k_{0}^{2}-i\delta]^{-1}
\nonumber\\ && \times
            [A^{\prime2}-p_{0}^{\prime2}-i\delta]^{-1}
            [A''+p_{0}^{\prime\prime}-i\delta]^{-1}
            [A^{2}-p_{0}^{2}-i\delta]^{-1}.
\end{eqnarray}
The energy-variable integrations can be performed in a straight forward
manner using the residue theorem, e.g.
\begin{equation}
  \int_{-\infty}^{+\infty}\!dk_{0}\ \frac{e^{i\beta k_{0}}}
    {\omega^{2}-k_{0}^{2}-i\delta} = \frac{2\pi i}{2\omega}
    e^{\mp i\beta\omega},
\end{equation}
where in the exponential the (--)-sign and the (+)-sign apply to
$\beta>0$ and $\beta<0$, respectively. Also,
\begin{equation}
  \int_{-\infty}^{+\infty}\!dp''_{0}\ \frac{e^{i\beta p''_{0}}}
    {A''+p''_{0}-i\delta} = \left\{ \begin{array}{lcl}
     2\pi i\ e^{-i\beta A''} & , & \beta > 0 \\
     0 & , & \beta< 0 \end{array} \right. .
\end{equation}
Keeping track of the signs in the exponentials, the intermediate
result is
\begin{eqnarray}
  {\cal J}_{1P} = (2\pi)^{-4} (2\pi i)^{5}
   \left[4 \omega \omega'\right]^{-1}\, &&
    \left\{ \int_{0}^{\infty}\!\!d\alpha\int_{\alpha}^{\infty}\!\!d\beta\
    e^{-i\alpha(\omega'+A'- A'')}
    e^{-i\beta(\omega+A+A'')}   \right.                  \nonumber\\
  && + \int_{-\infty}^{0}\!\!d\alpha \int_{0}^{\infty}\!\!d\beta\
    e^{+i\alpha(\omega'+A'+A'')}
    e^{-i\beta(\omega+A+A'')}                            \nonumber\\
  && \left. + \int_{-\infty}^{0}\!\!d\alpha\int_{\alpha}^{0}\!\!d\beta\
    e^{+i\alpha(\omega'+A'+A'')}
    e^{+i\beta(\omega+A-A'')} \right\}.
\end{eqnarray}
Performing the remaining elementary integrals, we end up with
\begin{eqnarray}
  {\cal J}_{1P} = -\frac{2\pi i}{4\omega\omega'}&&
     \left\{\frac{1}{A'+A''+\omega'}\,\frac{1}{A+A'+\omega+\omega'}\
\right.\nonumber\\ && \left.
    +\ \frac{1}{A'+A''+\omega'}\,\frac{1}{A+A''+\omega}\
    +\ \frac{1}{A+A''+\omega}\,\frac{1}{A+A'+\omega+\omega'} \right\}.
           \nonumber\\
  & & \label{I1bpair}
\end{eqnarray}
The terms in the curly brackets correspond to the different time-ordered
graphs (a), (b), and (c) of Fig.~\ref{pap2fig3}.

(ii) {\it The two-pair graph.}
Here the integral to be performed is
\begin{eqnarray}
  {\cal J}_{2P}({\bf p}',{\bf p}|W) &=& -(2\pi)^{-2}
      \left[ W-2E({\bf p}')\right]\left[W-2E({\bf p})\right]
\nonumber\\ && \times
      \int\!dp'_{0}\int\!dp_{0}\int\!dk'_{0}\int\!dk_{0}\
        \delta^{4}(p'-p-k-k')                          \nonumber\\
  && \times [k^{\prime 2}-m^{\prime 2}+i\delta]^{-1}\,
          \left[F^{(a)}_{W}({\bf p}',p'_{0})^{-1}\
          F^{(b)}_{W}(-{\bf p}',-p'_{0})\right]^{-1}   \nonumber\\
  && \times [k^{2}-m^{2}+i\delta]^{-1}\,
          \left[F^{(a)}_{W}({\bf p},p_{0})\
          F^{(b)}_{W}(-{\bf p},-p_{0})\right]^{-1}.
\end{eqnarray}
This integral can be brought into the form
\begin{eqnarray}
  {\cal J}_{2P} = -(2\pi)^{-3}  [4A'A] &&
    \int_{-\infty}^{+\infty}\!\!d\alpha
    \int_{-\infty}^{+\infty}\!\!dp'_{0}
    \int_{-\infty}^{+\infty}\!\!dp_{0}
    \int_{-\infty}^{+\infty}\!\!dk'_{0}
    \int_{-\infty}^{+\infty}\!\!dk_{0}\
    e^{i\alpha(p'_{0}-p_{0}-k_{0}-k'_{0})}                 \nonumber\\
  && \times [\omega'^{2}-k_{0}^{\prime 2}-i\delta]^{-1}
            [\omega^{2}-k_{0}^{2}-i\delta]^{-1}
\nonumber\\ && \times
            [A^{\prime2}-p_{0}^{\prime2}-i\delta]^{-1}
            [A^{2}-p_{0}^{2}-i\delta]^{-1}.
\end{eqnarray}
The evaluation of the energy-variable integrals gives in this case
\begin{equation}
  {\cal J}_{2P} = -(2\pi)^{-3} (2\pi i)^{4}
     \left[4 \omega \omega' \right]^{-1}\, \left\{
       \int_{0}^{\infty}\!d\alpha\ e^{-i\alpha(\omega+\omega'+A+A')}
     + \int_{-\infty}^{0}\!d\alpha\ e^{+i\alpha(\omega+\omega'+A+A')}
   \right\}.
\end{equation}
The two terms correspond to the time-ordered graph (d) of
Fig.~\ref{pap2fig3} and its ``mirror'' graph. Carrying out the final
$\alpha$ integrations leads to the final expression
\begin{equation}
  {\cal J}_{2P} = \frac{2\pi i}{4\omega\omega'}\,
                  \frac{2}{A+A'+\omega+\omega'}.    \label{I2pair}
\end{equation}

%\appendix
\section{ISOSPIN FACTORS}
\label{app:appB}
In this Appendix we review the calculation of the isospin factors.
The isospin factor comes from the evaluation of all possible
contractions for the vacuum matrix element of the product of the
meson fields coming from the contractions of the interaction Hamiltonians
at the different vertices.
For the one-pair diagrams we label the $N\!Nm_{1}m_{2}$ vertex with (1),
the meson vertex on the other nucleon line with (2), and the pion vertex
with (3). The pion is propagating with momentum ${\bf k}_{1}$ and the
other meson with momentum ${\bf k}_{2}$. The isospin factors are then
found to be
\begin{eqnarray*}
  (\pi\pi)_{0}  &:\ & \langle0| \bbox{\pi}\!\cdot\!\bbox{\pi}(1)
        \bbox{\tau}_{2}\!\cdot\!\bbox{\pi}(2)
        \bbox{\tau}_{2}\!\cdot\!\bbox{\pi}(3) |0\rangle =
        (\delta_{im}\delta_{in}+\delta_{in}\delta_{im})
        \tau_{2m}\tau_{2n} = 6,                                 \\
  (\sigma\sigma)&:\ & \langle0|\sigma\sigma(1)\sigma(2)\sigma(3)
        |0\rangle = (12,13+13,12)_{\rm contractions} = 2,       \\
  (\pi\pi)_{1}  &:\ & \langle0| \bbox{\tau}_{1}\!\cdot\!\bbox{\pi}
        \times\partial^{\mu}\bbox{\pi}(1)
        \bbox{\tau}_{2}\!\cdot\!\bbox{\pi}(2)
        \bbox{\tau}_{2}\!\cdot\!\bbox{\pi}(3)
        |0\rangle = \varepsilon_{ijk}\tau_{1k}\tau_{2m}\tau_{2n}
        \langle0| \pi_{i}(1)\partial^{\mu}\pi_{j}(1)\pi_{m}(2)\pi_{n}(3)
        |0\rangle                                               \\
        && \hspace*{5.6cm} = \varepsilon_{ijk}\tau_{1k}\tau_{2m}\tau_{2n}
        [\delta_{im}\delta_{jn}\partial^{\mu}(3)+\delta_{in}\delta_{jm}
        \partial^{\mu}(2)]                                     \\
        && \hspace*{5.6cm} = 2i\bbox{\tau}_{1}\!\cdot\!\bbox{\tau}_{2}
        (\partial_{1}^{\mu}-\partial_{2}^{\mu}),\\
  (\pi\rho)_{1} &:\ & \langle0| \bbox{\tau}_{1}\!\cdot\!\bbox{\pi}
        \times\bbox{\rho}(1) \bbox{\tau}_{2}\!\cdot\!\bbox{\rho}(2)
        \bbox{\tau}_{2}\!\cdot\!\bbox{\pi}(3)
        |0\rangle = \varepsilon_{ijk}\tau_{1k}\tau_{2m}\tau_{2n}
        \delta_{in}\delta_{jm}=-2i\bbox{\tau}_{1}\!\cdot\!\bbox{\tau}_{2},\\
  (\pi\sigma)   &:\ & \langle0|
        \bbox{\tau}_{1}\!\cdot\!\bbox{\pi}(1)
        \sigma(1)\sigma(2)\bbox{\tau}_{2}\!\cdot\!\bbox{\pi}(3)
        |0\rangle = \tau_{1i}\delta_{im}\tau_{2m} =
        \bbox{\tau}_{1}\!\cdot\!\bbox{\tau}_{2}.
\end{eqnarray*}
For the two-pair diagrams, we label the $N\!Nm_{1}m_{2}$ vertices on
each nucleon line by (1) and (2), respectively. We then find
\begin{eqnarray*}
  (\pi\pi)_{0}  &:\ & \langle0| \bbox{\pi}\!\cdot\!\bbox{\pi}(1)
       \bbox{\pi}\!\cdot\!\bbox{\pi}(2) |0\rangle =
       (\delta_{im}\delta_{im}+\delta_{im}\delta_{im}) = 6,        \\
  (\sigma\sigma)&:\ & \langle0| \sigma\sigma(1)\sigma\sigma(2)
       |0\rangle = 2,                                              \\
  (\pi\pi)_{1}  &:\ & \langle0| \bbox{\tau}_{1}\!\cdot\!\bbox{\pi}
       \times\partial^{\mu}\bbox{\pi}(1)
       \bbox{\tau}_{2}\!\cdot\!\bbox{\pi}
       \times\partial^{\nu}\bbox{\pi}(2) |0\rangle =
       \varepsilon_{ijk}\varepsilon_{mnl}\tau_{1k}\tau_{2l}
       \langle0| \pi_{i}(1)\partial^{\mu}\pi_{j}(1)\pi_{m}(2)
                 \partial^{\nu}\pi_{n}(2) |0\rangle                 \\
       && \hspace*{5.5cm} = 2\bbox{\tau}_{1}\!\cdot\!\bbox{\tau}_{2}
       (\partial_{2}^{\mu}\partial_{2}^{\nu}-
       \partial_{1}^{\mu}\partial_{2}^{\nu})                       \\
       && \hspace*{5.5cm} = \bbox{\tau}_{1}\!\cdot\!\bbox{\tau}_{2}
       (\partial_{1}^{\mu}-\partial_{2}^{\mu})
       (\partial_{1}^{\nu}-\partial_{2}^{\nu}),                    \\
  (\pi\rho)_{1} &:\ & \langle0| \bbox{\tau}_{1}\!\cdot\!\bbox{\pi}
       \times\bbox{\rho}(1) \bbox{\tau}_{2}\!\cdot\!\bbox{\pi}
       \times\bbox{\rho}(2) |0\rangle =
       \varepsilon_{ijk}\varepsilon_{mnl}\tau_{1k}\tau_{2l}\delta_{im}
       \delta_{jn} = 2\bbox{\tau}_{1}\!\cdot\!\bbox{\tau}_{2},     \\
  (\pi\sigma)   &:\ & \langle0| \bbox{\tau}_{1}\!\cdot\!\bbox{\pi}(1)
       \sigma(1)\bbox{\tau}_{2}\!\cdot\!\bbox{\pi}(2)\sigma(2)
       |0\rangle = \tau_{1i}\delta_{im}\tau_{2m} =
       \bbox{\tau}_{1}\!\cdot\!\bbox{\tau}_{2}.
\end{eqnarray*}
We should point out that for the $(\pi\pi)_{1}$ vertices the isospin
factors are defined to be $2i\bbox{\tau}_{1}\!\cdot\!\bbox{\tau}_{2}$ and
$\bbox{\tau}_{1}\!\cdot\!\bbox{\tau}_{2}$ for the one-pair and two-pair
potentials, respectively, because the $\partial^{\mu}_{i}$ dependence,
i.e., the ${\bf k}_{i}$ dependence, is already included in the operators
$O^{(n)}_{\alpha\beta,p}$ given in Tables~\ref{O1pair} and \ref{O2pair}.

%\appendix
\section{COORDINATE-SPACE POTENTIALS}
\label{app:appC}
The explicit formulas for the soft-core meson-pair potentials
are given below.

(i) Scalar $\pi\otimes\pi\ (I=0)$ pairs:
\begin{eqnarray}
  V^{(1)}_{\rm pair}(r) &=& 6 \frac{g_{(\pi\pi)_{0}}}{m_{\pi}}
        \frac{f_{N\!N\pi}^{2}}{m_{\pi}^{2}}
        \left[I'_{2,\pi}(r)\right]^{2},                  \label{tmep1}\\
  V^{(2)}_{\rm pair}(r) &=& -3 \frac{g_{(\pi\pi)_{0}}^{2}}{m_{\pi}^{2}}
    \frac{2}{\pi} \int_{0}^{\infty}\!d\lambda
    \left[F_{\pi}(\lambda,r)\right]^{2}.                   \label{tmep2}
\end{eqnarray}

(ii) Scalar $\sigma\otimes\sigma$ pairs:
\begin{eqnarray}
  V^{(1)}_{\rm pair}(r) &=& 2\frac{g_{(\sigma\sigma)}}{m_{\pi}}
          g_{N\!N\sigma}^{2} \left[I_{2,\sigma}(r)\right]^{2},
                                                        \label{tmep3}\\
  V^{(2)}_{\rm pair}(r) &=& -\frac{g_{(\sigma\sigma)}^{2}}{m_{\pi}^{2}}
         \frac{2}{\pi} \int_{0}^{\infty}\!d\lambda
         \left[F_{\sigma}(\lambda,r)\right]^{2}.           \label{tmep4}
\end{eqnarray}

(iii) Vector $\pi\otimes\pi\ (I=1)$ pairs:
\begin{eqnarray}
  V^{(1)}_{\rm pair}(r) &=& 4
    \left(\bbox{\tau}_{1}\!\cdot\!\bbox{\tau}_{2}\right)
    \frac{g_{(\pi\pi)_{1}}}{m_{\pi}^{2}}
    \frac{f_{N\!N\pi}^{2}}{m_{\pi}^{2}}
    \Biggl\{ \frac{2}{\pi}\int_{0}^{\infty}\!d\lambda
    \left[F'_{\pi}(\lambda,r)\right]^{2} - \frac{1}{M}
    \frac{1}{r^{2}}[I'_{2,\pi}(r)]^{2}\,{\bf L}\!\cdot\!{\bf S} \nonumber\\
  &&  - \frac{1+\kappa_{1}}{3M} \left[
    \frac{1}{r}I'_{2,\pi}\left(\frac{1}{r}I'_{2,\pi}
    +2I''_{2,\pi}\right)(r)
    \left(\bbox{\sigma}_{1}\!\cdot\!\bbox{\sigma}_{2}\right)
    +\frac{1}{r}I'_{2,\pi}
    \left(\frac{1}{r}I'_{2,\pi}-I''_{2,\pi}\right)(r)\
    S_{12} \right] \Biggr\},                                \label{tmep7}\\
  V^{(2)}_{\rm pair}(r) &=& -
    \left(\bbox{\tau}_{1}\!\cdot\!\bbox{\tau}_{2}\right)
    \frac{g_{(\pi\pi)_{1}}^{2}}{m_{\pi}^{4}}
    \frac{2}{\pi} \int_{0}^{\infty}\!d\lambda\ F_{\pi}(\lambda,r)
    \biggl[ \frac{\Lambda_{\pi}^{3}}{8\pi\sqrt{\pi}}\
    e^{-\frac{1}{4}\Lambda_{\pi}^{2}r^{2}}-2\lambda^{2}F_{\pi}
    (\lambda,r)\biggr].                                     \label{tmep8}
\end{eqnarray}

(iv) Axial $\pi\otimes\rho\ (I=1)$ pairs:
\begin{eqnarray}
  V^{(1)}_{\rm pair}(r) &=& -{\textstyle\frac{2}{3}}
     \left(\bbox{\tau}_{1}\!\cdot\!\bbox{\tau}_{2}\right)
     \frac{g_{(\pi\rho)_{1}}}{m_{\pi}}
     \frac{f_{N\!N\pi}}{m_{\pi}} \frac{g_{N\!N\rho}}{M_{N}}
     \Biggl\{ \left(1+\kappa_{\rho}\right) I'_{2,\pi}I'_{2,\rho}(r)
     \left[S_{12} - 2(\bbox{\sigma}_{1}\!\cdot\!\bbox{\sigma}_{2})\right]
                                          \nonumber\\
  && \hspace*{2.0cm} +\left(I''_{2,\pi}+\frac{2}{r} I'_{2,\pi}\right)
    I_{2,\rho}(r)\left(\bbox{\sigma}_{1}\!\cdot\!\bbox{\sigma}_{2}\right)
    + \left(I''_{2,\pi}-\frac{1}{r} I'_{2,\pi}\right)I_{2,\rho}(r)\
    S_{12} \Biggr\},                                       \label{tmep9}\\
  V^{(2)}_{\rm pair}(r) &=& -(\bbox{\tau}_{1}\!\cdot\!\bbox{\tau}_{2})
    (\bbox{\sigma}_{1}\!\cdot\!\bbox{\sigma}_{2})
    \frac{g_{(\pi\rho)_{1}}^{2}}{m_{\pi}^{2}}
    \frac{2}{\pi}\int_{0}^{\infty}\!d\lambda\ F_{\pi}F_{\rho}(\lambda,r).
                                                          \label{tmep10}
\end{eqnarray}

(v) Pseudovector $\pi\otimes\sigma$ pairs:
\begin{eqnarray}
  V^{(1)}_{\rm pair}(r) &=& {\textstyle\frac{2}{3}}
    \left(\bbox{\tau}_{1}\!\cdot\!\bbox{\tau}_{2}\right) g_{N\!N\sigma}
    \frac{g_{(\pi\sigma)}}{m_{\pi}^{2}} \frac{f_{N\!N\pi}}{m_{\pi}}
    \Biggl\{ \left[ \left(I''_{2,\pi}+\frac{2}{r}I'_{2,\pi}
    \right) I_{2,\sigma} - I'_{2,\pi} I'_{2,\sigma} \right](r)
    \left(\bbox{\sigma}_{1}\!\cdot\!\bbox{\sigma}_{2}\right)  \nonumber\\
  && \hspace*{3.0cm}+ \left[ \left(I''_{2,\pi}-\frac{1}{r} I'_{2,\pi}\right)
    I_{2,\sigma} - I'_{2,\pi} I'_{2,\sigma} \right](r)\
    S_{12} \Biggr\},                                     \label{tmep11}\\
  V^{(2)}_{\rm pair}(r) &=& {\textstyle\frac{1}{6}}
    \left(\bbox{\tau}_{1}\!\cdot\!\bbox{\tau}_{2}\right)
    \frac{g_{(\pi\sigma)}^{2}}{m_{\pi}^{4}}
    \frac{2}{\pi}\int_{0}^{\infty}\!d\lambda \Biggl\{ \left[
    \left(F''_{\pi}+\frac{2}{r} F'_{\pi}\right) F_{\sigma} +
    \left(F''_{\sigma}+\frac{2}{r} F'_{\sigma}\right) F_{\pi}
    -2 F'_{\pi} F'_{\sigma} \right](\lambda,r)
    \left(\bbox{\sigma}_{1}\!\cdot\!\bbox{\sigma}_{2}\right)  \nonumber\\
  && \hspace*{2.8cm} +\left[ \left(F''_{\pi}-\frac{1}{r} F'_{\pi}\right)
    F_{\sigma} + \left(F''_{\sigma}-\frac{1}{r} F'_{\sigma}\right) F_{\pi}
   -2F'_{\pi}F'_{\sigma}\right](\lambda,r)\ S_{12}\Biggr\}.\label{tmep12}
\end{eqnarray}

\narrowtext

\widetext
\begin{table}
\caption{The one-pair isospin factors $C^{(1)}(\alpha\beta)$ and
         momentum operators $O^{(1)}_{\alpha\beta,p}({\bf k}_{1},\omega_{1};
         {\bf k}_{2},\omega_{2})$. The index $p$ labels the three
         time-ordered contributions of Fig.~\protect\ref{pap2fig3} and
         is only of relevance for the $(\pi\pi)_{1}$ entry, where they
         are shown as a row vector.
         Note that $\kappa_{1}=(f/g)_{(\pi\pi)_{1}}$.}
\begin{tabular}{lll}
 $(\alpha\beta)$ & $C^{(1)}(\alpha\beta)$ &
 $O_{\alpha\beta,p}^{(1)}({\bf k}_{1},\omega_{1};{\bf k}_{2},\omega_{2})$\\
\tableline
 $(\pi\pi)_{0}$  & $6$                                          &
      $-{\bf k}_{1}\!\cdot{\bf k}_{2}
      +{\textstyle\frac{i}{2}}(\bbox{\sigma}_{1}+\bbox{\sigma}_{2})
       \!\cdot\!({\bf k}_{1}\times{\bf k}_{2}) $             \\[0.2cm]
 $(\sigma\sigma)$& $2$                                &  $1$  \\[0.2cm]
 $(\pi\pi)_{1}$  & $2i\bbox{\tau}_{1}\!\cdot\!\bbox{\tau}_{2}$ &
     $i\left[{\bf k}_{1}\!\cdot\!{\bf k}_{2}
       -{\textstyle\frac{i}{2}}(\bbox{\sigma}_{1}+\bbox{\sigma}_{2})
       \!\cdot\!({\bf k}_{1}\times{\bf k}_{2})\right]
           \left(\omega_{1}-\omega_{2};\ -\omega_{1}-\omega_{2};\
           -\omega_{1}+\omega_{2}\right) $                   \\[0.2cm]
 & & $ \hspace{2ex} + {\displaystyle\frac{i}{M}} \left[ (1+\kappa_{1})
      \bbox{\sigma}_{1}\!\cdot\!({\bf k}_{1}\times{\bf k}_{2})
      \bbox{\sigma}_{2}\!\cdot\!({\bf k}_{1}\times{\bf k}_{2})
     +{\textstyle\frac{i}{2}}(\bbox{\sigma}_{1}+\bbox{\sigma}_{2})
       \!\cdot\!({\bf k}_{1}\times{\bf k}_{2})\,{\bf q}\!\cdot\!
                ({\bf k}_{1}-{\bf k}_{2})\right]$            \\[0.2cm]
 $(\pi\rho)_{1}$ & $-2i\bbox{\tau}_{1}\!\cdot\!\bbox{\tau}_{2}$ &
     ${\displaystyle\frac{i}{M}}
     \left[\bbox{\sigma}_{1}\!\cdot\!{\bf k}_{1}
           \bbox{\sigma}_{2}\!\cdot\!{\bf k}_{1}
       +{\textstyle\frac{1}{2}}(1+\kappa_{\rho})
        (\bbox{\sigma}_{1}\!\cdot\!{\bf k}_{1}
         \bbox{\sigma}_{2}\!\cdot\!{\bf k}_{2} +
         \bbox{\sigma}_{1}\!\cdot\!{\bf k}_{2}
         \bbox{\sigma}_{2}\!\cdot\!{\bf k}_{1} -
        2\bbox{\sigma}_{1}\!\cdot\!\bbox{\sigma}_{2}
        {\bf k}_{1}\!\cdot\!{\bf k}_{2}) \right]$           \\[0.2cm]
 $(\pi\sigma)$   & $\bbox{\tau}_{1}\!\cdot\!\bbox{\tau}_{2}$    &
     $\left[\bbox{\sigma}_{1}\!\cdot\!{\bf k}_{1}
            \bbox{\sigma}_{2}\!\cdot\!{\bf k}_{2} +
            \bbox{\sigma}_{1}\!\cdot\!{\bf k}_{2}
            \bbox{\sigma}_{2}\!\cdot\!{\bf k}_{1} -
           2\bbox{\sigma}_{1}\!\cdot\!{\bf k}_{1}
            \bbox{\sigma}_{2}\!\cdot\!{\bf k}_{1} \right] $
\end{tabular}
\label{O1pair}
\end{table}

\narrowtext
\begin{table}
\caption{The two-pair isospin factors $C^{(2)}(\alpha\beta)$ and
         momentum operators $O^{(2)}_{\alpha\beta,p}({\bf k}_{1},\omega_{1};
         {\bf k}_{2},\omega_{2})$. The index $p$ labels the two
         time-ordered contributions of Fig.~\protect\ref{pap2fig3} but
         is of no relevance in the final result.}
\begin{tabular}{lll}
 $(\alpha\beta)$ & $C^{(2)}(\alpha\beta)$ &
 $O_{\alpha\beta,p}^{(2)}({\bf k}_{1},\omega_{1};{\bf k}_{2},\omega_{2})$\\
\tableline
 $(\pi\pi)_{0}$  & $6$                                   & $1$ \\[0.2cm]
 $(\sigma\sigma)$& $2$                                   & $1$ \\[0.2cm]
 $(\pi\pi)_{1}$  & $\bbox{\tau}_{1}\!\cdot\!\bbox{\tau}_{2}$    &
     $(\omega_{1}-\omega_{2})^{2}$                             \\[0.2cm]
 $(\pi\rho)_{1}$ & $2\bbox{\tau}_{1}\!\cdot\!\bbox{\tau}_{2}$   &
     $\bbox{\sigma}_{1}\!\cdot\!\bbox{\sigma}_{2}$             \\[0.2cm]
 $(\pi\sigma)$   & $\bbox{\tau}_{1}\!\cdot\!\bbox{\tau}_{2}$    &
     $\bbox{\sigma}_{1}\!\cdot\!({\bf k}_{1}-{\bf k}_{2})
      \bbox{\sigma}_{2}\!\cdot\!({\bf k}_{1}-{\bf k}_{2})$
\end{tabular}
\label{O2pair}
\end{table}

\narrowtext
\begin{table}
\caption{Pair-meson coupling constants employed in the potentials shown
         in Figs.~\protect\ref{pap2fig5} to \protect\ref{pap2fig7}.
         Coupling constants are at ${\bf k}^{2}=0$.
         All coupling constants are fixed at their theoretical
         values as derived in Sec.~\protect\ref{sec:chap5}.}
\begin{tabular}{cldd}
 $J^{PC}$ & $(\alpha\beta)$  & $g/4\pi$  & $f/4\pi$ \\
\tableline
 $0^{++}$ & $(\pi\pi)_{0}$   &--0.412  &        \\
 $0^{++}$ & $(\sigma\sigma)$ &--0.482  &        \\
 $1^{--}$ & $(\pi\pi)_{1}$   &  0.058  & 0.216  \\
 $1^{++}$ & $(\pi\rho)_{1}$  &  0.598  &        \\
 $1^{++}$ & $(\pi\sigma)$    &  0.053  &
\end{tabular}
\label{gpair}
\end{table}

\mediumtext
\begin{table}
\caption{$\chi^{2}$ and $\chi^{2}$ per datum ($\chi^{2}_{\rm p.d.p.}$)
         at the 10 energy bins for the updated partial-wave analysis
         (PWA) and the ESC and Nijm93 potential models. $N_{\rm data}$
         lists the number of data within each energy bin. The bottom
         line gives the results for the total 0--350 MeV interval.}
\begin{tabular}{r@{--}lrrrrrrr}
    \multicolumn{2}{c}{} &  & \multicolumn{2}{c}{PWA}
  & \multicolumn{2}{c}{ESC} & \multicolumn{2}{c}{Nijm93} \\
  \multicolumn{2}{c}{Bin(MeV)} & $N_{\rm data}$ & $\chi^{2}$
     & $\chi^{2}_{\rm p.d.p.}$ & $\chi^{2}$ & $\chi^{2}_{\rm p.d.p.}$
     & $\chi^{2}$ & $\chi^{2}_{\rm p.d.p.}$ \\
\tableline
 0.0&0.5 &  145 &  144.45 & 0.996 &  146.15 & 1.008 &  185.78 & 1.28 \\
 0.5&2   &   68 &   42.97 & 0.632 &   47.38 & 0.697 &   55.06 & 0.81 \\
   2&8   &  110 &  106.28 & 0.966 &  115.19 & 1.047 &  128.92 & 1.17 \\
   8&17  &  296 &  276.31 & 0.933 &  326.31 & 1.102 &  368.20 & 1.24 \\
  17&35  &  359 &  279.54 & 0.829 &  332.10 & 0.925 &  393.92 & 1.10 \\
  35&75  &  585 &  567.18 & 0.970 &  697.19 & 1.192 & 1337.28 & 2.29 \\
  75&125 &  399 &  409.58 & 1.027 &  421.89 & 1.057 &  480.96 & 1.20 \\
 125&183 &  760 &  820.69 & 1.080 &  936.53 & 1.232 & 1443.10 & 1.90 \\
 183&290 & 1047 & 1035.48 & 0.989 & 1261.50 & 1.205 & 1995.71 & 1.91 \\
 290&350 &  992 &  997.02 & 1.005 & 1706.75 & 1.721 & 2866.39 & 2.89 \\
\tableline
   0&350 & 4761 & 4697.50 & 0.987 & 5990.99 & 1.258 & 9255.32 & 1.94
\end{tabular}
\label{tabchi2}
\end{table}

\narrowtext
\begin{figure}
\caption{Feynman one-pair and two-pair diagrams.}
\label{pap2fig1}
\end{figure}

\begin{figure}
\caption{``Duality'' picture of the meson-pair potentials.
         $R_{i}$ represents a nucleon resonance, $H_{j}$ the
         intermediating heavy boson, and $M_{1}$ and $M_{2}$
         are the mesons being exchanged.}
\label{pap2fig2}
\end{figure}

\begin{figure}
\caption{Time-ordered (a--c) one-pair and (d) two-pair diagrams.
         The dotted line with momentum ${\bf k}_{1}$ refers to the
         pion and the dashed line with momentum ${\bf k}_{2}$ refers
         to one of the other (vector, scalar, or pseudoscalar) mesons.
         To these we have to add the ``mirror'' diagrams, where
         for the one-pair diagrams the pair vertex occurs on
         the other nucleon line.}
\label{pap2fig3}
\end{figure}

\begin{figure}
\caption{Meson saturation of a pair vertex. The mesons are denoted
         by $M_{1}$ and $M_{2}$, and the intermediating heavy boson
         by $H$.}
\label{pap2fig4}
\end{figure}

\begin{figure}
\caption{Central, spin-spin, and tensor components of the $I=0$
         one-pair and two-pair potentials for (a) $r\leq1$ fm and
         (b) $1\leq r\leq2$ fm. The pairs considered are $0^{++}$:
         $(\pi\pi)_{0}$ and $(\sigma\sigma)$;
         $1^{--}$: $(\pi\pi)_{1}$; and
         $1^{++}$: $(\pi\rho)_{1}$ and $(\pi\sigma)$.}
\label{pap2fig5}
\end{figure}

\begin{figure}
\caption{Same as Fig.~\protect\ref{pap2fig5}, but for $I=1$.}
\label{pap2fig6}
\end{figure}

\begin{figure}
\caption{Spin-orbit contributions of the one-pair $0^{++}$ and $1^{--}$
         potentials for both $I=0$ and $I=1$. A distinction is made
         between the nonadiabatic and pseudovector-vertex contributions.}
\label{pap2fig7}
\end{figure}

\end{document}